\documentclass{kapproc} 

\usepackage{procps} 

\usepackage[dvips]{graphicx}

\upperandlowercase

 \let\footnote\savefootnote

\let\footnotetext\savefootnotetext

\kluwerbib 

\begin{document}

\articletitle[]
{The Evolution of Cluster Substructure with Redshift}

\author{Tesla E. Jeltema\altaffilmark{1}, Claude R. Canizares\altaffilmark{1}, Mark W. Bautz\altaffilmark{1}, and David A. Buote\altaffilmark{2}}

\affil{\altaffilmark{1}Massachusetts Institute of Technology, \ 
\altaffilmark{2}University of California at Irvine}
\email{tesla@space.mit.edu}


\anxx{T. E. Jeltema et al.}

\begin{abstract}

Using Chandra archival data, we quantify the evolution of cluster morphology with redshift.  To quantify cluster morphology, we use the power ratio method developed by Buote and Tsai (1995).  Power ratios are constructed from moments of the two-dimensional gravitational potential and are, therefore, related to a cluster's dynamical state.  Our sample will include 40 clusters from the Chandra archive with redshifts between 0.11 and 0.89.  These clusters were selected from two fairly complete flux-limited X-ray surveys (the ROSAT Bright Cluster Sample and the Einstein Medium Sensitivity Survey), and additional high-redshift clusters were selected from recent ROSAT flux-limited surveys.  Here we present preliminary results from the first 28 clusters in this sample.  Of these, 16 have redshifts below 0.5, and 12 have redshifts above 0.5.

\end{abstract}

\section{Introduction and Sample Selection}

Clusters form and grow through mergers with other clusters and groups.  Substructure or a disturbed cluster morphology indicates that a cluster is dynamically young (i.e. it will take some time for it to reach a relaxed state), and the amount of substructure in clusters in the present epoch and how quickly it evolves with redshift depend on the underlying cosmology.  In low density universes, clusters form earlier and will be on average more relaxed in the present epoch.  Clusters at high redshift, closer to the epoch of cluster formation, should be on average dynamically younger and show more structure.  In addition, the evolution of cluster morphology is important to the understanding of many cluster properties including mass, gas mass fraction, lensing properties, and galaxy morphology and evolution.

Several studies have been done to quantify substructure in clusters at low redshift (e.g., Jones \& Forman 1992; Mohr et al. 1995; Buote \& Tsai 1996).  However, it is only with recent X-ray and optical surveys that we are beginning to find tens of clusters with z $>$ 0.8, and it is becoming possible to study the evolution of substructure.  Using the power ratio method (Buote \& Tsai 1995), we are studying structure in a sample of 40 clusters observed with the Chandra X-ray Observatory.  As a first cut, our sample includes only clusters with a redshift above 0.1 so that a reasonable area of each cluster will fit on a Chandra CCD.  In order to have a reasonably unbiased sample, clusters were selected from the BCS (Ebeling et al. 1998) and EMSS (Gioia \& Luppino 1994) surveys.  They were also required to have a luminosity greater than $5 \times 10^{44}$ ergs s$^{-1}$, as listed in those catalogs.  Additional high-redshift clusters were selected from recent ROSAT flux-limited surveys (Rosati et al. 1998; Perlman et al. 2002; Gioia et al. 2003; Vikhlinin et al. 1998).  This led to a sample of 40 clusters with redshifts between 0.11 and 0.89.  Here we present the results from 28 of these clusters.  Sixteen of these have redshifts below 0.5 with an average redshift of 0.26; the other twelve have redshifts above 0.5 and an average redshift of 0.72.

\section{Power Ratios}

Power ratios are constructed from moments of the two-dimensional gravitational potential.  They are capable of distinguishing many cluster morphologies, and they have been shown to distinguish different cosmological models (Buote \& Xu 1997; Valdarnini, Ghizzardi, \& Bonometto 1999; Suwa et al. 2003).  Essentially, this method involves calculating multipole moments of the X-ray surface brightness.  The moments, $a_m$ and $b_m$, are given below.  Here $\Sigma$ is the surface brightness.  These are calculated in a circle of radius R centered on the centroid of emission.
\begin{eqnarray}
a_m(R) & = & \int_{R^{\prime}\le R} \Sigma(\vec x^{\prime})
\left(R^{\prime}\right)^m \cos m\phi^{\prime} d^2x^{\prime} \nonumber \\
b_m(R) & = & \int_{R^{\prime}\le R} \Sigma(\vec x^{\prime})
\left(R^{\prime}\right)^m \sin m\phi^{\prime} d^2x^{\prime} \nonumber
\end{eqnarray}

The moments are sensitive to asymmetries in the surface brightness distribution and are, therefore, sensitive to substructure.  The powers (shown below) are the sum of the squares of the moments, and the power ratios are formed by dividing by $P_0$ to normalize out flux.  The physical motivation for the power ratio method is that it is based on the multipole expansion of the two-dimensional gravitational potential ($\Psi$).  With $\Sigma$ as the surface mass density, the powers are the squares of the multipole moments of $\Psi$ evaluated over a circle of a given radius.  Below $\Psi^{\rm int}_m$ is the {\itshape m\/}th multipole of the 2D gravitational potential due to matter interior to the circle of radius R, and $\langle\cdots\rangle$ represents the azimuthal average around the circle.  Ignoring factors of 2G, the powers are

\begin{equation}
P_0=\left[a_0\ln\left(R\right)\right]^2=\langle(\Psi^{\rm int}_0)^2\rangle, \nonumber
\end{equation}
\begin{equation}
P_m={1\over 2m^2 R^{2m}}\left( a^2_m + b^2_m\right)=\langle(\Psi^{\rm int}_m)^2\rangle. \nonumber
\end{equation}

In the case of X-ray studies, X-ray surface brightness replaces surface mass density in the calculation of power ratios.  X-ray surface brightness is proportional to gas density squared and generally shows the same qualitative structure as the projected mass density, allowing a similar quantitative classification of clusters.

\section{Preliminary Results}

For each cluster in our initial sample of 28, we calculated P$_2$/P$_0$, P$_3$/P$_0$, and P$_4$/P$_0$ centered on the cluster centroid (where P$_1$ vanishes).  We use an aperture radius of 0.5 Mpc.  At larger radii, the high-z clusters have low S/N, and the low-z clusters become too large to fit on a Chandra CCD.  Figure 1 shows a plot of P$_2$/P$_0$ versus P$_3$/P$_0$.  High-redshift clusters (z$>$0.5) are plotted with diamonds and have red error bars.  Low-redshift clusters are plotted with asterisks and have blue error bars.  The error bars were found using Monte Carlo simulations and represent 90\% confidence.  The different power ratios are sensitive to different types of structure and looking at correlations among them can help distinguish cluster morphologies.  In these plots, the most disturbed clusters appear at the upper-right, and the most relaxed clusters at the lower-left.

The high-z and low-z cluster samples have similar P$_2$/P$_0$ ratios, but the high-z clusters tend to have higher P$_3$/P$_0$.  One possible reason that P$_3$/P$_0$ is better at distinguishing the high-redshift clusters from the low-redshift ones is that it is not sensitive to ellipticity: a purely elliptical cluster will only have even multipoles.  Large odd multipoles unambiguously indicate asymmetry (substructure) in a cluster.  P$_4$/P$_0$, which is similar to P$_2$/P$_0$ but sensitive to smaller scale structure, also appears to distinguish the two samples, especially when correlated with P$_3$/P$_0$.  Figure 1 also shows a plot of P$_3$/P$_0$ versus P$_4$/P$_0$.  Here the high-z clusters tend to be in the upper corner (more structure) and the low-z clusters tend to be in the bottom corner (less structure).  A Mann-Whitney rank-sum test shows that the P$_3$/P$_0$ and P$_4$/P$_0$ ratios for the high-redshift clusters are on average larger than those for the low-redshift clusters at 95\% significance.

At this point, we have analyzed 28 out of 40 clusters in our sample, and it appears that the amount of structure in clusters increases with redshift.  Specifically, P$_3$/P$_0$ and P$_4$/P$_0$ are higher for the high-redshift clusters.  These clusters were all selected to have high luminosities, and we do not believe the difference in power ratios is due to a difference in luminosities between the two samples.  Using a radius of a fixed over-density rather than a fixed physical size also does not account for the difference.  These and other possible systematic effects will be addressed in more detail later.  In addition to completing the analysis of our Chandra sample, we also plan to compare the sample to numerically simulated clusters provided by Greg Bryan.

\begin{figure}
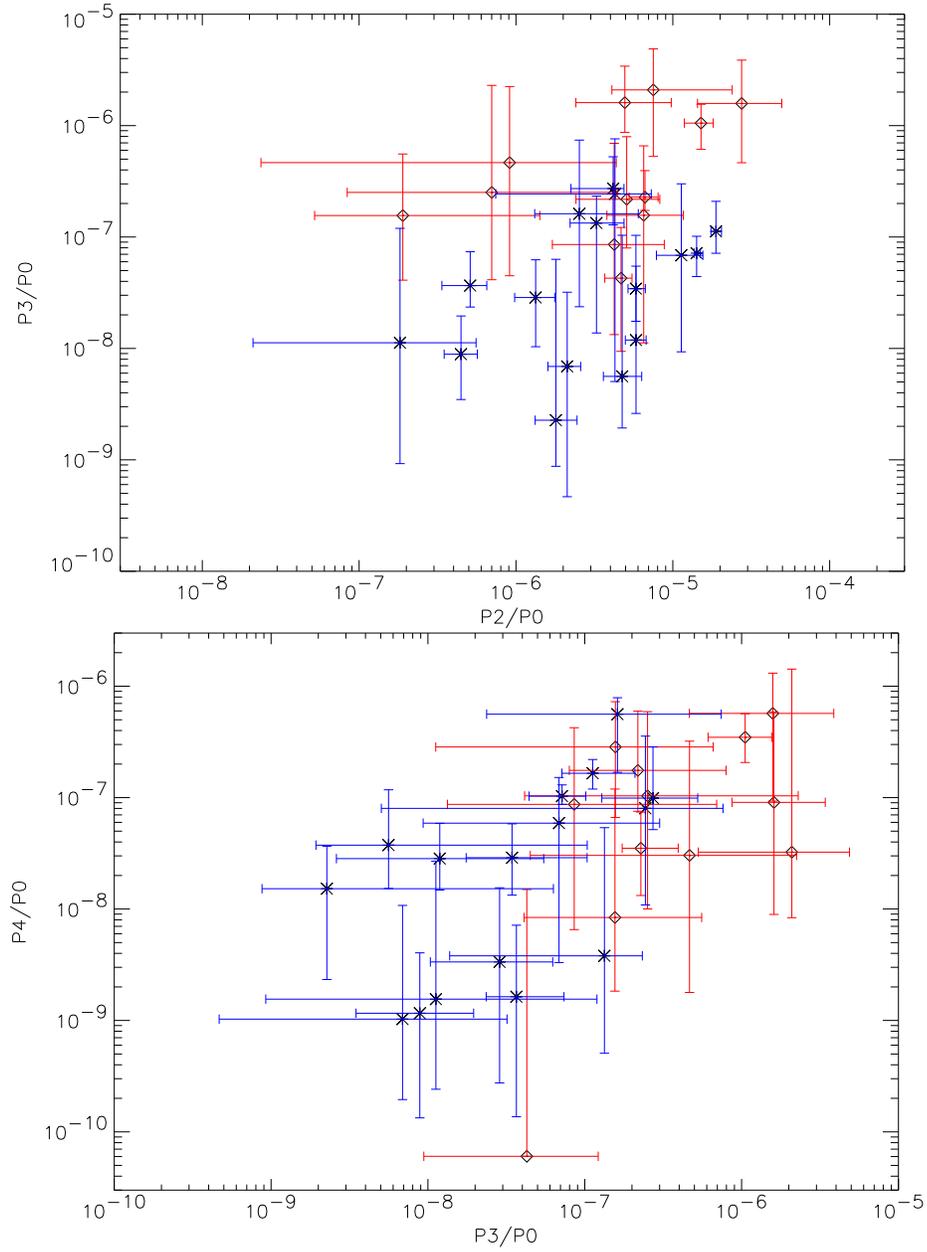

\centering
\includegraphics[scale=0.7,angle=0]{p2vsp3_err.epsi}
\includegraphics[scale=0.7,angle=0]{p3vsp4_err.epsi}
\caption{Power ratios computed in a 0.5 Mpc aperture for the first 28 clusters in our sample.  High-redshift clusters (z$>$0.5) are plotted with diamonds and have red error bars.  Low-redshift clusters are plotted with asterisks and have blue error bars.  Top: P$_2$/P$_0$ versus P$_3$/P$_0$.  Bottom: P$_3$/P$_0$ versus P$_4$/P$_0$.}
\end{figure}

\end{document}